\documentclass[
    onecolumn, preprintnumbers,
    amsmath, amssymb,
    prd,
    showpacs, showkeys
]{revtex4}
\usepackage{graphicx}%
\usepackage{amsmath}
\usepackage[normalem,normalbf]{ulem}
\usepackage{amssymb}
\usepackage{bm}
\usepackage{enumerate}

\newcommand{\nn}{\nonumber}
\begin{document}
\title{$\kappa-$Minkowski spacetime and a uniformly accelerating observer}
\author{Hyeong-Chan Kim}
\email{hckim@phya.yonsei.ac.kr}
\author{Jae Hyung Yee}%
\email{jhyee@phya.yonsei.ac.kr}
\affiliation{Deptartment of Physics, Yonsei University,
Seoul 120-749, Republic of Korea.
}%
\author{Chaiho Rim}
\email{rim@chonbuk.ac.kr}
\affiliation{Department of Physics and Research Institute
of Physics and Chemistry, Chonbuk National University,
Chonju 561-756, Korea.
}%
\date{\today}%
\bigskip
\begin{abstract}
\bigskip

We analyze the response of a detector with a uniform
acceleration $\alpha$ in $\kappa-$Minkowski spacetime using
the first order perturbation theory. The monopole detector
is coupled to a massless complex scalar field in such a way
that it is sensitive to the Lorentz violation due to the
noncommutativity of spacetime present in the
$\kappa-$deformation. The response function deviates from
the thermal distribution of Unruh temperature at the order
of $1/\kappa$ and vanishes exponentially as the proper time
of the detector exceeds a certain critical time, a
logarithmic function of $\kappa$. This suggests that the
Unruh temperature becomes not only fuzzy but also
eventually decreases to zero in this model.

\end{abstract}
\maketitle

\section{Introduction}

In the last decade, there has been a great interest in
attempting to explain the cosmic observational data as a
quantum gravitational effect. As a theoretical framework to
study the quantum gravity effects phenomenologically,
several field theories on noncommutative
spacetimes~\cite{wess,giovanni,girelli,okon,freidel,amelino,livine,amelino2}
have been studied.

In this direction of research, one common aspect is
the introduction of deformed symmetries and results in Lorentz
symmetry breaking.
This is reformulated in noncommutative spacetimes
where a dimensional parameter is introduced, related to
the Planck mass.
This dimensional parameter is expected to suppress the
Lorentz symmetry breaking
in the commutative limit. However, careful estimates
suggest that there may exist a strong  fine-tuning problem
in noncommutative spacetime approach at a one-loop level \cite{finetuning}.
In addition, unitarity of the theory  is in question in
noncommutative spacetime field theories \cite{unitarity1,unitarity2}.
This consideration, however,  does not decrease
the interest in noncommutative spacetime field theories.
On the contrary, one needs to find a good candidate
for a realistic model.

The $\kappa$-Minkowski spacetime introduces a dimensionful
deformation parameter $\kappa$, whose natural choice is to
put $\kappa=M_P$, the Planck mass
\cite{kosinski,glikman,daszkie}. The $\kappa-$Minkowski
spacetime respects rotational symmetry and appears to be a
good candidate to study the quantum gravity effect. Scalar
field theory has been studied by introducing the
differential structure  in $\kappa-$Minkowski
space~\cite{kosinski,freidel2}. The $\kappa-$deformation is
extended to the curved space with $\kappa-$Robertson-Walker
metric and is applied to cosmic microwave background
radiation in~\cite{kim2}.

On the other hand, it is an interesting question how the quantum gravity effect changes the structures of vacua and particle in
curved spacetime.
Unruh~\cite{unruh,birrel} calculated the response of a
particle detector moving with a uniform acceleration,
under the assumption that the state of field is initially
in its vacuum state, {\it i.e.}, the Minkowski vacuum
and the field interacts only with the detector.
It is shown that the detector responds as if it would have remained un-accelerated but immersed in a heat bath at
temperature {\it acceleration}/$2\pi$, the Unruh temperature.
This is called the Unruh effect and has been studied further in~\cite{wald}.
The idea has been extended to include back-reaction
problem and generalized to realistic
detectors~\cite{massar,kim}.

The Unruh effect is interesting since
it gives an analogy of the Hawking radiation in
blackhole spacetime due to the thermal form of the
transition probability in the presence of event horizon. In
this point of view, the accelerating frame (Rindler
spacetime) is regarded as a simplest toy model simulating
the radiation effect of a blackhole.
Since it is not easy to calculate the
field theory in curved-noncommutative spacetime containing
a blackhole directly, one may instead, seek for the effect
on an accelerating frame first.

In the present paper, we study the response function
of a uniformly accelerating monopole particle detector
which interacts with massless complex scalar field in
$\kappa-$Minkowski spacetime.
Since the primary  effect of $\kappa-$deformation
is the Lorentz symmetry breaking, we design a
particle detector model so that it is sensitive to this
symmetry violation.

In Sec. II, we briefly summarize the complex scalar field
theory in $\kappa-$Minkowski spacetime using the
differential structure given in ~\cite{kosinski}. We derive
the Feynman propagator from the action and propose Wightman
functions. In Sec. III, we consider a uniformly
accelerating detector interacting with massless complex
scalar field and calculate the transition amplitude of the
detector. It is found that the Lorentz symmetry violation
effect appears to the response function at $O(1/\kappa)$.
We summarize the results in Sec. IV.

\section{Scalar field theory in $\kappa$-Minkowski spacetime}

In this section, we briefly introduce
the $\kappa-$deformation
of the Minkowski spacetime
and construct the scalar field theory. Some details can be
found in Ref.~\cite{kosinski}. Here we use the signature
of spacetime metric, $(+,-,-,-)$.

\subsection{$\kappa-$Minkowski spacetime}

The time and space coordinates are not commuting in
$\kappa$-Minkowski spacetime but satisfy the commutation
relations
\begin{equation} \label{k-Minkowski}
~[\hat x^0, \hat x^i] = \frac{i}{\kappa} \hat x^i\,,\qquad
~[\hat x^i, \hat  x^j]=0\,,\qquad i,j=1,2,3\,.
\end{equation}
$\kappa$ is a positive parameter which represents the
deformation of the space-time. Its Hopf algebra $H_x$ is
described  by the co-product
\begin{equation*} 
~\Delta(\hat x^\mu)= \hat x^\mu \otimes 1 + 1 \otimes \hat x^\mu \,.
\end{equation*}
The $\kappa$-deformed Poincar\'e algebra is constructed
using commuting four-momenta
$[p^\mu, p^\nu]=0$ and the dual Hopf algebra $H_p$.
The co-product of four momenta is given as
\begin{equation} \label{co-product:p}
~ \Delta(p^0)=p^0\otimes 1+1\otimes p^0\,, \qquad
~\Delta(p^i)=p^i\otimes e^{-p^0/\kappa}+1\otimes p^i\,.
\end{equation}

Exponential operator $e^{-i p \cdot \hat x}$ is the basic
ingredient which transforms the theory in space-time
coordinates to the theory in momentum space. Here $\hat x
=(\hat x^0, \hat {\textbf{x}})$, $p=(p^0,\textbf{p})$ and
$p\cdot \hat x\equiv p^0 \hat x^0 - {\bf p \cdot \hat x}$.
Its ordering is defined as
\begin{eqnarray} \label{NOrdering}
:e^{-i p \cdot \hat x}:\,
\equiv e^{-i p^0 \hat x^0}\, e^{i \bf p \cdot \hat x}\,.
\end{eqnarray}
Multiplication of two ordered exponentials follows from
Eqs.~(\ref{k-Minkowski}) and (\ref{NOrdering}):
\begin{eqnarray} 
  :e^{-i p \hat x}::e^{-i q \hat x}:= :e^{i ({\bf p}
     e^{-q^0/\kappa}+ {\bf q}){\bf \hat x}-i (p^0+q^0)
      \hat x^0}: ,
\end{eqnarray}
which is indicated in the four momentum addition rule
described by the coproduct~(\ref{co-product:p}).

To find the differential calculus one differentiates
Eq.~(\ref{k-Minkowski}) and finds
\begin{eqnarray*} 
~[\tau^0,\hat x^i]+[\hat x^0,\tau^i]=\frac{i}{\kappa}
    \tau^i,\qquad
~[\tau^i,\hat x^j]+[\tau^j,\hat x^i]= 0\, ,
\end{eqnarray*}
where $\tau^\mu= d\hat x^\mu$ is the exterior derivatives
along the four dimensional space-time direction. One may
choose the commutation relations
\begin{eqnarray}\label{alg1}
~[\tau^0,\hat x^i]=\frac{i}{\kappa}\tau^i,~~~~
~[\tau^i,\hat x^0] =0 ,
\end{eqnarray}
and fix the unique differential
structure through Jacobi's identities:
\begin{eqnarray} \label{diff}
~[\tau^\mu, \hat x^\nu]=
\frac{1}{4}\eta^{\mu\nu}\tau^5-\frac{i}{\kappa}\eta^{\mu\nu}
\tau^0 +\frac{i}{\kappa} \eta^{0\mu} \tau^\nu ,~~~~
[\tau^5,\hat x^\mu]= -\frac{4}{\kappa^2}\tau^\mu \,.
\end{eqnarray}
This demonstrates that
the differential calculus is not closed in 4-dimensions
but needs a new exterior derivative along the fifth direction
$\tau^5$~\cite{sitarz,gonera}. One may identify $\tau^5$ as
\begin{equation*}
\tau^5=[\tau^\mu,x_\mu]+\frac{3i}{\kappa} \tau^0\,.
\end{equation*}
The covariance property under the action of
$\kappa-$deformed Poincar\'{e} group was established in
Ref.~\cite{kosinski,daszkie} and a five-dimensional
bi-covariant differential calculus was considered in
Ref.~\cite{kosinski}.

Using $\tau^A$ with  $A=0,1,2,3,5$, one writes the
derivative of the exponential function as
\begin{eqnarray} \label{chis} 
d:e^{-ip\hat x}:\,&=& -i : \chi_A(p)e^{-ip\hat x}: \tau^A
\\ \nn
\chi_0(p) &=& \kappa\left[\sinh
\frac{p_0}{\kappa}+\frac{{\bf p}^2}{2\kappa^2}
e^{p_0/\kappa}\right]
\\ \nn
\chi_i(p)&=& p_i e^{p_0/\kappa}
\\ \nn
\chi_5(p)&=&-\frac{i}{8}M_\kappa^2(p) \,.
\end{eqnarray}
$M_\kappa^2(p)$ is the first Casimir operator of the
algebra (\ref{co-product:p}),
\begin{eqnarray} \label{Casimir}
M_\kappa^2(p)=\left(2 \kappa \sinh \frac{p_0}{2 \kappa
}\right)^2- {\bf p}^2 e^{p_0/\kappa}\,,
\end{eqnarray}
which is invariant under the deformed Poincar\'e
transformations.
As $\kappa \rightarrow \infty$, this reduces to the
commutative relation, $M_\infty^2(p)=p_0^2-{\bf p}^2=m^2$.
For finite $\kappa$, the on-shell three-momentum is bounded
from above by $ {\bf p}^2 \leq \kappa^2$~\cite{kow} since
\begin{eqnarray*}
\frac{{\bf p}^2}{\kappa^2}=1
-\left(2+\frac{M_\kappa^2}{\kappa^2}\right)e^{-p_0/\kappa}+
    e^{-2p_0/\kappa}
\end{eqnarray*}
and $M_\kappa^2$ is a non-negative constant.
$p^0$ goes to infinity
when the momentum reaches the maximal value $\kappa$.
In addition, defining  5-dimensional momentum
\[
{\cal P}_A = (\chi_0\,, \vec{\chi} \,,
\kappa+4i \chi_5/\kappa  )\,,
\]
we realize that the momentum space constitutes a
4-dimensional de Sitter spacetime:
\[
({\cal P}_0)^2 - ({\cal P}_i)^2 - ({\cal P}_5)^2
 = - \kappa^2 \,.
 \]

\subsection{Scalar field representation}

Scalar field in $\kappa$-Minkowski spacetime is represented
using the $\kappa$-deformed Fourier transformation. Suppose
$\tilde\Phi(p)$ is a classical function in commuting
four-momentum space. The scalar field on the
$\kappa$-Minkowski spacetime is then defined as
\begin{eqnarray} \label{fourier}
\Phi(\hat x) = \int_p \tilde \Phi(p)\,  :e^{-i p \hat x}:
\end{eqnarray}
where $\int_p $ denotes $\displaystyle \int
\frac{d^4p}{(2\pi)^4}\,$. This scalar field on the
non-commuting coordinates smoothly reduces to the commuting
case if $\kappa \to \infty$. The definition also allows us
to transform the integration over non-commuting coordinates
unambiguously into the ones with commuting momentum space
integration if one uses the delta-function relation
\begin{equation}
(2\pi)^4 \,\delta^4 (p)= \int_{\hat x} :e^{-i p \hat x}:
\end{equation}
where $\int_{\hat x} =  \int d ^4 \hat x $.
For example, the multiplication of fields integrated over $\hat x$
is expressed as
\begin{eqnarray}\label{phi^2}
\int_{\hat x}  \Phi^2(\hat x)  = \int_p
\tilde\Phi(p)\, \tilde{\Phi}(-p^0, -{\bf p} e^{p^0/\kappa}) \,.
\end{eqnarray}

One may obtain a conjugate field if one defines the
conjugation of ordered exponential as
\begin{equation} \label{exp-dagger}
\Big(:e^{-ip \hat x}:\Big)^\dagger
= e^{-i \textbf{p} \cdot \hat{\textbf{x} }} \,
  e^{i p^0 \hat x^0}
= :e^{ -i( e^{p_0/\kappa} ) \textbf{p}\cdot \hat{\textbf{x}}
    + i p^0 \hat x^0} : \,.
\end{equation}
With this, we have
\begin{eqnarray}
\Big( \Phi^\dagger (\hat x) \Big)^\dagger = \Phi(\hat x)\,,
\qquad \Big( \Phi_1 (\hat x) \Phi_2 (\hat x) \Big)^\dagger
=  \Phi_2^\dagger (\hat x) \Phi_1^\dagger (\hat x),
\end{eqnarray}
and  the conjugate scalar field is given as
\begin{eqnarray} \label{conju-field}
\Phi^{\dagger}(\hat x)&=&
\int_p \tilde \Phi(p)^\dagger \,
\Big(:e^{-ip \cdot \hat x}:\Big)^\dagger
\equiv \int_p \tilde \Phi^{c} (p) \, :e^{-ip \cdot \hat x}:
\\
\tilde \Phi^c (p)
&=& e^{3p_0/\kappa}\,
    \tilde \Phi^\dagger(-p_0, -e^{p_0/\kappa} {\bf p} )\,.
    \nn
\end{eqnarray}
$\Phi^\dagger(p)$ represents the ordinary complex conjugate
of $\Phi(p)$ for classical field and hermitian conjugate
for quantum field.  Likewise in Eq.~(\ref{phi^2}), one may
write the multiplication of fields as
\begin{eqnarray} \label{12dagger}
\int_{\hat x} \Phi_1^\dagger (\hat x) \Phi_2 (\hat x)
&=&  \int_p  \tilde \Phi_1^\dagger (p)\, \tilde \Phi_2 (p)
\\
\int_{\hat x} \Big(\Phi_1 (\hat x) \Phi_2 (\hat x)
\Big)^\dagger &=& \int_{\hat x}  \Phi_2^\dagger(\hat x)
\Phi_1^\dagger (\hat x) = \int_p  \tilde \Phi_2^c (-{\bf
p}e^{p_0/\kappa},-p_0) \tilde \Phi_1^c (p) .\nn
\end{eqnarray}

If the scalar field $\Phi(\hat x)$ is  real, we need
$\Phi(\hat x) ^\dagger = \Phi(\hat x) $. This, in turn
gives $\tilde \Phi^c (p)= \tilde \Phi(p)$. (Our definition
differs from the one in Ref.~\cite{kosinski}. There,
$\tilde \Phi^c (p) $ is replaced by $\tilde \Phi ^\dagger(p)$.
)

The partial derivative of field is defined from the partial
derivative of the exponential functions in
Eq.~(\ref{chis}),
\begin{eqnarray} \label{derivative}
\hat \partial_\mu   \, :e^{-ip \cdot \hat x}: \,
= -i \, \chi_\mu  (p) \, :e^{-ip \cdot \hat x}:  \,.
\end{eqnarray}
The adjoint derivative $\hat \partial_\mu ^\dagger$
is obtained from the relation
\begin{eqnarray}
\int_{\hat x} \Phi_1(\hat x) \,
\hat \partial_\mu \, \Phi_2(\hat x)
= \int _{\hat x} \Big( \hat\partial_\mu ^\dagger\,
\Phi_1(\hat x)\Big)\,  \Phi_2(\hat x) \nn
\end{eqnarray}
which leads to
\begin{eqnarray}
\hat \partial_\mu^\dagger \, \, :e^{-ip \cdot \hat x}:
&=& -i  \chi_\mu ^\dagger(p) \, :e^{-ip \cdot \hat x}:
\\ \chi_\mu ^\dagger(p)&=&
\chi_\mu  (-p^0, -{\bf p} e^{p_0/\kappa} )\,.
\nn
\end{eqnarray}
This results in a useful relation
\begin{equation}
\Big( \hat \partial_\mu  \Phi (\hat x) \Big)^\dagger
= - \hat \partial_\mu ^\dagger \, \Phi^\dagger  (\hat x)  \,.
\end{equation}

\subsection{Free field action of a complex field}
The free field action of a  complex scalar field in
$\kappa-$Minkowski spacetime can be written in analogy with
the commutative one as
\begin{eqnarray} \label{Sx:0}
S= \int_{\hat x}
 \left[(\hat \partial_\mu^\dagger \Phi^\dagger(\hat x)) \,
 \hat \partial^\mu\Phi(\hat x)
  - m^2 \Phi^\dagger(\hat x)\, \Phi(\hat x)\right].
\end{eqnarray}
Here the fifth direction is omitted and evaluated in 4-dimensions
only.  This action has not  $SO(4,1)$ invariance in \cite{freidel2}
but the deformed Poincar\'e symmetry is respected. (See Eq.~(\ref{Sp:0})
below).
This action can be written in momentum space representation
by using the Fourier transform~(\ref{fourier}),
\begin{eqnarray} \label{Sp:0}
S &=& \int_p \tilde \Phi^\dagger(p)\,
\Delta_F^{-1}(p)\, \tilde\Phi(p)
\\
\Delta_F ^{-1}(p)&=& \left[
    M^2_\kappa(p)\left(1+\frac{M^2_\kappa(p)}{
    4\kappa^2}\right)-m^2+i\epsilon\right]\nn \,,
\end{eqnarray}
where $\epsilon$ is the usual small number prescription for
the Feynman propagator. This action shows that the
non-commutative spacetime modifies the propagator for the
free fields defined in the commutative coordinate space
\begin{equation}
\phi(x) = \int_p e^{-ip\cdot x }\, \tilde \Phi(p) \,.
\end{equation}

The number of poles in $\Delta_F$ are, however, infinitely
many on the complex momentum plane. Explicit pole positions
are given as $\omega_n ^\pm =\omega^{\pm}+ i n \kappa\pi$
with $n$ an arbitrary integer;
\begin{equation}
 \omega^{\pm}= -\kappa\ln
\left( \sqrt{1+ \frac{m^2-i\epsilon}{\kappa^2}}
\mp \sqrt{ \frac{m^2+\textbf{p}^2 -i\epsilon}{\kappa^2}}\right)\,.
\end{equation}
For the massless case, $\omega^{\pm}$ reduces to
\begin{eqnarray}\label{mass-shell:0}
 \omega^\pm = -\kappa \ln
    (1\mp \frac {|\textbf{p}|-i\epsilon }{\kappa})\,.
\end{eqnarray}
$\omega^\pm$ represent the two stable on-shell spectra and
due to the restriction $|\textbf{p}^2 | < \kappa^2$,
$\omega^+ $ is positive  and  $\omega^- $ is negative. The
existence of the two stable spectra suggests that one may
define the Minkowski vacuum $|0_M\rangle$ which reduces to
commutative vacuum when $\kappa \to \infty$. Then the
time-ordered product is given as
\begin{eqnarray}
G_F(x,y)  =
\langle 0_M| T \phi(x) \phi^\dagger(y) |0_M\rangle
    = \int_p  e^{-ip\cdot (x-y)}i\Delta_F(p)=G_F(x-y) \,,
\end{eqnarray}
while
\begin{eqnarray*}
\langle 0_M| T \phi(x) \phi(y) |0_M\rangle
    &=&\langle 0_M| T \phi^\dagger(x) \phi^\dagger(y) |0_M\rangle
= 0 \,.
\end{eqnarray*}
The asymmetry of the pole positions of the Feynman
propagator in Eq.~(\ref{Sp:0}), $ \omega^+ + \omega^- =
-\kappa\ln ( 1- \textbf{p}^2/\kappa^2 ) \ne 0$,
demonstrates that the Lorentz symmetry is broken at the
order of $1/\kappa$. It is natural to define $\omega^+$ as
the particle spectrum and $\omega^-$ as the anti-particle
one. In this case, the vacuum $|0_M\rangle $ does not
respect the particle and antiparticle symmetry.

In addition, it is noted that when $\Delta x^0 >0$, $G_F(\Delta x) $
creates not only the excitation with $\omega^+$ but also
unstable ones with $\omega^+_n$ and  $\omega^-_n$ with $n$
negative integers. Thus we may define particle spectra as
$\omega^+$, $\omega^+_n$ and $\omega^-_n$ with $n<0$.
Similarly, when $\Delta x^0 <0$, $G_F(\Delta x ) $ creates the stable
excitation with $\omega^-$ as well as unstable ones with
$\omega^+_n$ and $\omega^-_n$ with $n$ positive integers,
which indicates that  $\omega^-$, $\omega^+_n$ and
$\omega^-_n$ with $n>0$ represent anti-particle spectra.

The positive Wightman function $W_+(x, y)$ is defined as
\begin{eqnarray}
W_+(x,y) &=& \langle 0_M| \phi(x) \phi^\dagger(y)
|0_M\rangle \,.
\end{eqnarray}
$W_+(x, y)$ measures the amplitude to create particles including the
unstable ones and is defined to be the Feynman propagator when
$\Delta x^0>0$.
Thus it can be represented as
\begin{eqnarray} \label{pwight}
W_+(x,y) =W_+(x-y)=\sum_{\omega =\omega_+, \omega_{n<0}^{\pm} }
 \int\frac{d^3{\bf p}}{(2\pi)^3 2|{\bf p}|}
    \frac{e^{-i \omega (x_0 -y_0)
    +i {\bf p\cdot (x-y)}}}{ 1
    +M_\kappa^2(\omega,{\bf p})/(2\kappa^2) }\,.
\end{eqnarray}
This result can be written formally as
\begin{eqnarray*}
W_+(\Delta x )=\int_{p+}   e^{-i p\cdot \Delta x}
2\pi \,\delta \left( M_\kappa^2 \left(1+
\frac{M_\kappa^2}{4\kappa^2}\right) -m^2\right)\,.
\end{eqnarray*}
Here $\int_{p+}$ indicates that the integral over $p^0$ includes
not only the real mass-shell position $\omega^+$
but also the complex ones, $\omega^+_n$ and $\omega^-_n$ with $n<0$.
The delta function integration is evaluated using the property
\begin{eqnarray} \label{dM}
\left.\frac{\partial M_\kappa^2}{\partial
p_0}\right|_{\omega^\pm}=\left[
    \kappa\left(1-\frac{\textbf{p}^2}{\kappa^2}\right)e^{p_0/\kappa}
    -\kappa e^{-p_0/\kappa}
    \right]_ {\omega^\pm}= \pm 2|\textbf{p}|\,.
\end{eqnarray}

Likewise, the negative Wightman function $W_-(x, y)$ is
defined as
\begin{eqnarray} \label{nwight}
W_-(x,y) &=& \langle 0_M| \phi^\dagger(y) \phi(x) |0_M\rangle
 =W_-(x-y)\\
&=& \sum_{\omega =\omega_-, \omega_{n>0}^{\pm} }
 \int\frac{d^3{\bf p}}{(2\pi)^3 2|{\bf p}|}
    \frac{e^{-i \omega (x_0 -y_0)
    +i {\bf p\cdot (x-y)}}}{ 1
    +M_\kappa^2(\omega,{\bf p})/(2\kappa^2) }
    \, \nn \,.
\end{eqnarray}

The explicit form of $W_\pm (\Delta x)$ for
massless case is needed in the next section.
When $\kappa |\Delta x^0|\gg 1$,
the complex mass-shell contributions  decay exponentially,
representing the unstable excitations.
As a result, among the infinitely many contributions to
$W_\pm (\Delta x)$, the main contribution comes from the
$\omega^\pm $ part. After the angular integrations
and rescaling $|\textbf{p}|$ by $\kappa$, we have
\begin{eqnarray} \label{W:G}
W_+(\Delta x)&=&   \frac{\kappa}{8 i\pi^2
|\Delta\textbf{x}|} (g_+(\Delta x)-g_-(\Delta x)) ,
\\ \nn
g_\pm (\Delta x) &=&
 \int_0^1 dz~
e^{i\kappa [\log(1-z) \, \Delta x_0 \pm \, z  |\Delta
\textbf{x}|]} \,.
\end{eqnarray}
Using the result in Appendix A, we have, to the order of
$1/\kappa$,
\begin{eqnarray} \label{diff:Gpm}
W_+(\Delta x) &=& -\frac{1}{4\pi^2 } \left\{\xi - \frac{
i\Delta x_0}{\kappa } \left(\frac{3(\Delta x_0)^2 + (\Delta
\textbf{x})^2)} {\xi} \right ) \right\}^{-1}
+O(\kappa^{-2}) \, ,
\end{eqnarray}
where  $\xi=(\Delta x_0)^2 - (\Delta \textbf{x})^2$.
For $W_-(\Delta x)$, we have
\begin{eqnarray}
W_-(\Delta x) &=&  \frac{\kappa}
{8 i\pi^2 |\Delta \textbf{x}|}
( h_+(\Delta x)-h_-(\Delta x))
\\
h_\pm (\Delta x) &=&
 \int_0^1 dz~
e^{i\kappa (\log(1+z) \, \Delta x_0 \pm \, z  |\Delta
\textbf{x}|)}\,. \nn
\end{eqnarray}
To the order of $1/\kappa$, we have
\begin{eqnarray} \label{diff:Gmm}
W_-(\Delta x) &=& -\frac{1}{4\pi^2 } \left\{\xi - \frac{
i\Delta x_0}{\kappa} \left(\frac{3(\Delta x_0)^2 + (\Delta
\textbf{x})^2)} {\xi} \right ) \right\}^{-1}
+O(\kappa^{-2}) .
\end{eqnarray}

Here we assume that $\kappa \Delta x_0 \gg 1$ and $\xi \gg
1$, even though  $W_\pm(\Delta x)$ holds for  $\xi \sim  0$
in the commutative limit. In addition, it
is noted that the Lorentz symmetry breaking is reflected in
this result since $W_-(-\Delta x) \ne W_+(\Delta x)$ at the
order of $1/\kappa$.

\section{particle detector interacting with a massless scalar field
in $\kappa-$Minkowski space}

Suppose a  particle detector is moving along a world line
$X^\mu(\tau)$, where $\tau$ is the detector's proper time.
Unruh~\cite{unruh} calculated the response of the particle
detector moving with a uniform acceleration: If the
detector interacts with a free massless field, and the
system initially lies in the Minkowski vacuum state, then
the detector responds as if it were immersed in a heat bath
in an un-accelerated frame whose temperature turns out to
be {\it acceleration}/$2\pi$, the Unruh temperature. This
effect is called the Unruh effect.

In the Minkowski spacetime, we will let the detector
interact with a massless complex scalar field through the
detector's monopole moments, ${\cal M}(\tau)$ and ${\cal
M}^\dagger(\tau)$. The interaction is written as
\begin{eqnarray} \label{Ac}
S_I&=&c \int d\tau \,\int_x \delta^4(x-X(\tau)) \Big({\cal
M}^\dagger(\tau)\, \phi(x) + \phi^\dagger(x){\cal M}(\tau)
\Big)
\\&=&
\int d\tau \,  \Big( {\cal M}^\dagger(\tau)\,
 \phi(X(\tau)) + \phi^\dagger(X(\tau)) {\cal M}(\tau)\Big)
\nn
\end{eqnarray}
where $c$ is a small coupling constant.

In the $\kappa$-Minkowski spacetime, the complex scalar
field is affected by the non-commutative nature of the
spacetime and its action is written as Eq.~(\ref{Sp:0})
with vanishing mass. For the detector part, on the other
hand, we assume that it experiences the time evolution
under the ordinary quantum mechanics while moving along the
commutative world line $X^\mu(\tau)$.

The detector can be
coupled to the field in various ways. To find the
possibility we consider a $\kappa$-deformed delta function
\begin{eqnarray} \label{delta}
\delta^{(4)}(\hat x- X(\tau))
= \int_p :e^{-ip (\hat x-X(\tau))}: \,
= \int_p  e^{i pX(\tau)}\, :e^{-ip \hat x}: \,.
\end{eqnarray}
This delta function gives the property
\begin{eqnarray}
\int_{\hat x}  \Phi^\dagger ( \hat x) \, \delta^{(4)} ( \hat x- X(\tau ))
=  \int_p \tilde \Phi^\dagger (p) e^{ip\cdot X(\tau)}
=\Phi^\dagger (X(\tau))\,.
\end{eqnarray}
However, the $\kappa$-deformed delta
function is not self-conjugate, since
\begin{eqnarray*}
\delta^{(4)\dagger}(\hat x-X(\tau))
=\int_p  e^{i p X(\tau)} \,
 \Big( :e^{-ip \cdot \hat x }:\Big)^\dagger
= \int_p  \,e^{3p_0/\kappa}
e^{i p_0 X_0(\tau)-ie^{p_0/\kappa}\vec p\cdot \vec X(\tau)}\,
:e^{-i p \hat x}:  &\ne&
\delta(\hat x-X(\tau)) \,.
\end{eqnarray*}
Thus we need
\begin{eqnarray}
\Phi (X(\tau))
=\int_{\hat x}  \delta^{(4)\dagger} ( \hat x- X(\tau )) \,
\Phi ( \hat x)
\,.
\end{eqnarray}
Thanks to this property of the delta function, we may
rewrite the interaction Eq.~(\ref{Ac}) for the interaction
in the $\kappa-$Minkowski spacetime as
\begin{eqnarray}\label{SI}
S_I &=&
 c \int d\tau\,\left[ {\cal M}^\dagger(\tau)\int_{\hat x}
\delta^\dagger(\hat x-X(\tau)) \,\Phi(\hat x) + {\cal
M}(\tau)\int_{\hat x}\Phi^\dagger(\hat x)\, \delta^4(\hat
x-X(\tau)) \right]\,.
\end{eqnarray}
This choice is the simplest one, in the sense that
complicated coproduct of the $\kappa-$Minkowski spacetime
does not appear in the interaction term.
There exist other choices of interaction,
which may give more complicated non-commutative effect but this possibility is not considered in the present paper.

Suppose the detector lies initially in its ground state
$|E_0\rangle$ of a quantum mechanical hamiltonian $H_0$. As
the detector moves along a trajectory, it will in general
find itself undergoing a transition from the initial to an
excited state $|E\rangle$ with $E> E_0$. In addition, the
field might be affected by the detector since the field is
coupled to the detector and there will be a field
contribution to the transition.

We assume for simplicity that the field initially is in the
ground state $|0_M\rangle$, the vacuum state with respect
to the Minkowski spacetime and finally makes a transition
to an excited state $|\phi\rangle$, which may be assumed to
be a one particle state. Then we expect the whole
transition amplitude of the system $A_{f\leftarrow i}$ to
be evaluated by the first order perturbation theory:
\begin{eqnarray} \label{T:0}
A_{f\leftarrow i}&=& i c\langle E, \phi|
    \int_{-\infty}^{\tau_0}
d\tau \left[ {\cal M}^\dagger (\tau)\, \int_p
    \tilde \Phi(p)e^{-i p X(\tau)}
    + {\cal M}(\tau)\, \int_p
    \tilde \Phi^\dagger (p)e^{i p X(\tau)}
    \right]
    |0_M,E_0\rangle \,
\end{eqnarray}
where $\tau_0$ is the time when the detector
and the field reach the final state $|\phi,E\rangle$.
Using the time evolution of the monopole moment operator ${\cal M}(\tau)$
\begin{eqnarray} 
{\cal M}(\tau)= e^{iH_0\tau} {\cal M}(0) e^{-i H_0\tau}\,,
\quad {\cal M}^\dagger(\tau) = e^{iH_0\tau} {\cal
M}^\dagger(0) e^{-i H_0\tau}\, ,
\end{eqnarray}
$A_{f\leftarrow i}$ may factorize to give
\begin{eqnarray} \label{pert:1}
A_{f\leftarrow i} &= & ic \langle
    E|{\cal M}^\dagger(0)|E_0\rangle
        \int_{-\infty}^{\tau_0} d\tau\,
    e^{i(E-E_0)\tau}\, \int_p\, e^{-i p X(\tau)}
  \langle \phi| \tilde \Phi(p) |0_M\rangle  \\
 &&   +ic \langle
    E|{\cal M}(0)|E_0\rangle
        \int_{-\infty}^{\tau_0} d\tau\,
    e^{i(E-E_0)\tau}\, \int_p\, e^{i p X(\tau)}
    \langle \phi|\tilde \Phi^\dagger (p)|0_M\rangle \,.\nn
\end{eqnarray}

We first consider the detector moving along the world line
with a constant velocity $\textbf v$, the inertial world
line,
\begin{eqnarray} \label{iner}
X^0=t=\gamma\,\tau\,,\qquad \textbf {X}(\tau ) =\textbf
{X}(0) + \textbf{v}\, \gamma\,\tau , \nn
\end{eqnarray}
where $\gamma = 1/\sqrt{1-\textbf{v}^2/c^2}$.
Then the integral in~(\ref{pert:1}) with $\tau_0
\rightarrow \infty$ becomes
\begin{eqnarray} \label{pert:inertial}
A_{f\leftarrow i} &=& i c \langle E|{\cal M}^\dagger (0)|E_0\rangle
    \int_p\,
    2\pi\delta \left( E-E_0- \gamma\, (p^0-
        \textbf{p} \cdot \textbf{v} )\right)\,
        e^{i \textbf{p} \cdot \textbf{x}}\,
        \langle \phi|\tilde \Phi(p)|0_M\rangle  \nn\\
   &&+i c \langle E|{\cal M}(0)|E_0\rangle
    \int_p\,  2\pi\delta \left( E-E_0+\gamma\, (p^0-
        \textbf{p} \cdot \textbf{v} )\right)\,
        e^{-i \textbf{p} \cdot \textbf{x}}\,
        \langle \phi|\tilde \Phi^\dagger(p)|0_M\rangle
       \nn . \nonumber
\end{eqnarray}
The expectation values $\langle
\phi|\tilde\Phi(p)|0_M\rangle$ ($\langle
\phi|\tilde\Phi^\dagger(p)|0_M\rangle$) is nonzero only for
the case of $p_0<0$ ($p_0>0$) as in the commutative case.
In addition, since $E>E_0$ and $|p_0| \geq  \textbf{p}
\cdot \textbf{v}$, the delta functions make $A_{f\leftarrow
i}$ vanish, which reflects the fact that the
$\kappa$-Minkowski spacetime has the time-translational
invariance.

To explore the case when the detector follows a uniformly
accelerating path with acceleration $\alpha$,
whose coordinates  are chosen as
\begin{eqnarray} \label{UApath}
 X^0(\tau)=\alpha^{-1} \sinh \alpha \tau\,, \qquad
 X^1(\tau)= \alpha^{-1}\cosh \alpha \tau\,, \qquad
 X^2(\tau)=0=X^3(\tau)\,,
\end{eqnarray}
we consider the transition probability,
\begin{eqnarray} \label{pb}
|A_{f\leftarrow i}|^2 = c^2\sum _E \Big( M_+(E,E_0) \,
{\cal F}_+(E-E_0) + M_-(E,E_0) \, {\cal F}_-(E-E_0)\Big) ,
\end{eqnarray}
where we sum over intermediate states using  the
completeness of the basis $\{|\phi\rangle\}$ and
\begin{eqnarray*}
M_\pm (E,E_0) = \frac{1}{2}\left(|\langle E|{\cal
M}^\dagger (0)|E_0\rangle |^2 \pm |\langle E|{\cal
M}(0)|E_0\rangle |^2 \right) \,.
\end{eqnarray*}
Note that $M_+(E,E_0)$ is always larger than
$|M_-(E,E_0)|$.
 ${\cal F}_\pm$ is the part due to the response of the
field and is given by
\begin{eqnarray*}
{\cal F}_\pm (E) &=&\int_{-\infty}^{\tau_0} d\tau
\int_{-\infty}^{\tau_0}d\tau'\, e^{-iE(\tau-\tau')}\, f_\pm
(X(\tau), X(\tau'))\,,
\end{eqnarray*}
where
\begin{eqnarray*}
f_\pm (X(\tau), X(\tau'))
&=& \langle 0_M|  \left[ \Phi(X(\tau)) \,\Phi^\dagger (X(\tau'))
\pm \Phi^\dagger (X(\tau)) \, \Phi(X(\tau')) \right]|0_M\rangle
\\ &=& W_+(X(\tau)- X(\tau'))
 \pm  W_-(X(\tau')- X(\tau)) = f_\pm (X(\tau)- X(\tau')) \,.
\end{eqnarray*}
In the Lorentz invariant systems such as in the commutative
field theory, ${\cal F}_- (E) $ vanishes since $W_+( \Delta
X ) = W_-( -\Delta X ) $. In our case, ${\cal F}_- (E) $
does appear at the order of $1/\kappa$.

Excitation rate $R(\tau_0,E)$ is the rate of the transition
probability,
\begin{eqnarray} \label{exrate}
R(\tau_0,E)\equiv \frac{d |A_{f\leftarrow
    i}|^2}{d\tau_0}
= c^2\,\sum _E \,\Big( M_+( E, E_0) \,S_+(\tau_0,E) + M_-
(E, E_0) \,S_- (\tau_0,E) \Big)\, ,
\end{eqnarray}
where $ S_\pm(\tau_0,E) $ is called the response
function~\cite{resp} representing the Unruh effect:
\begin{eqnarray} \label{ext:resp}
S_\pm(\tau_0,E)&\equiv &\frac{d {\cal F}_\pm(E)}{d\tau_0}
\\
&=&\int_{-\infty}^{\tau_0} d\tau\,
    e^{iE(\tau-\tau_0)} f_\pm(X(\tau_0),X(\tau))
 + \int_{-\infty}^{\tau_0} d\tau\,
    e^{-iE(\tau-\tau_0)} f_\pm(X(\tau),X(\tau_0) ) .
\nn
\end{eqnarray}
Shifting $\tau$ by $\tau_0$ we have
\begin{eqnarray} \label{S:t}
S_\pm(\tau_0,E) &=&\int_{-\infty}^0  d\tau~
    e^{iE\tau} \,\, f_\pm(X(\tau_0),X(\tau+\tau_0))
 + \int_{-\infty}^0  d\tau~
    e^{-iE\tau} \,\, f_\pm(X(\tau+\tau_0),X(\tau_0) )
\\ \nn
&=&\int_{-\infty}^{\infty}   d\tau~
    e^{-iE\tau} \,\,
f_\pm(X(\tau_0-|\tau|/2+\tau/2), X(\tau_0-|\tau|/2-\tau/2) ).
\end{eqnarray}
For the massless case, $f_\pm( \Delta X )$ is given by
\begin{eqnarray}
f_+ (\Delta X)
&=& -\frac{1}{2\pi^2 \xi }
 +O(\kappa^{-2})\, , \\
f_-(\Delta X)
&=& -\frac{ i}{2\pi^2 \kappa }
\frac{  \Delta X^0 \Big( 3(\Delta X^0)^2 + (\Delta
\textbf{X})^2 \Big)} {\xi^3} +O(\kappa^{-2})  \,.
\end{eqnarray}

Along the uniformly accelerating path in
Eq.~(\ref{UApath}), $\Delta X = X(\tau)- X(\tau') $ is
parameterized as  as
\begin{eqnarray}
\Delta X^0 &=& \frac2{\alpha} \sinh\Big(\frac{\alpha (
\tau-\tau') }2\Big) \cosh\Big(\frac{\alpha(\tau+\tau')}2
\Big) ,
\\
\Delta X^1  &=& \frac2{\alpha}
\sinh\Big(\frac{\alpha ( \tau-\tau') }2\Big) \,
\sinh\Big(\frac{\alpha(\tau+\tau')}2\Big)
 \nn \\
\xi &=& (\Delta X^0 )^2
    -(\Delta X^1 )^2 = \frac4 {\alpha^2}
\sinh^2\Big(\frac{\alpha ( \tau-\tau') }2\Big)\,,\nn
\end{eqnarray}
and we have
\begin{eqnarray}
f_+(\Delta X) &=&-\frac{\alpha^2}{8\pi^2 } \left\{ \frac
1{\sinh^2 \frac{\alpha(\tau-\tau')}{2}} \right\}
+O\left(\frac1{\kappa^2} \right) ,
\\ \nonumber
f_-(\Delta X) &=&-\frac{i \alpha^3}{8\pi^2\, \kappa }
\left\{ \frac{ 4 \cosh^3 \frac{\alpha(\tau+\tau')}{2} -
\cosh \frac{\alpha(\tau+\tau')}{2}} {\sinh^3
\frac{\alpha(\tau-\tau')}{2}} \right\}
+O\left(\frac1{\kappa^2} \right) \,.
\end{eqnarray}
Thus $f_\pm $ in Eq.~(\ref{S:t}) is given explicitly in
terms of  the proper-time parametrization
\begin{eqnarray} \label{W:Ttau}
f_+(X((\tau_0-|\tau|/2+\tau/2),
        X(\tau_0-|\tau|/2-\tau/2))
&=& -\frac{\alpha^2}{8\pi^2}
\frac 1{\sinh^2(\frac{\alpha\tau}{2})}
 +O(\kappa^{-2}), \nn
\\
f_-(X((\tau_0-|\tau|/2+\tau/2),
        X(\tau_0-|\tau|/2-\tau/2))
&=&- \frac{i\alpha^3}{8\pi^2 \kappa}
\left\{ \frac  { \cosh 3 \alpha\tau_0 \, \cosh
\frac{3\alpha\tau}2 + 2\cosh \alpha\tau_0 \, \cosh
\frac{\alpha\tau}2   }{
\sinh^3(\frac{\alpha\tau}{2}) } \right.
\nn \\
&&\qquad\qquad
- \epsilon(t) \left.
\frac {\sinh 3 \alpha\tau_0 \,(2
\cosh \alpha\tau + 1) + 2\sinh \alpha\tau_0 }{
\sinh^2(\frac{\alpha\tau}{2}) }
\right\}
 +O(\kappa^{-2})\,. \nn
\end{eqnarray}
\begin{figure}[htbp]
\begin{center}
\includegraphics[width=.5\linewidth,origin=tl]{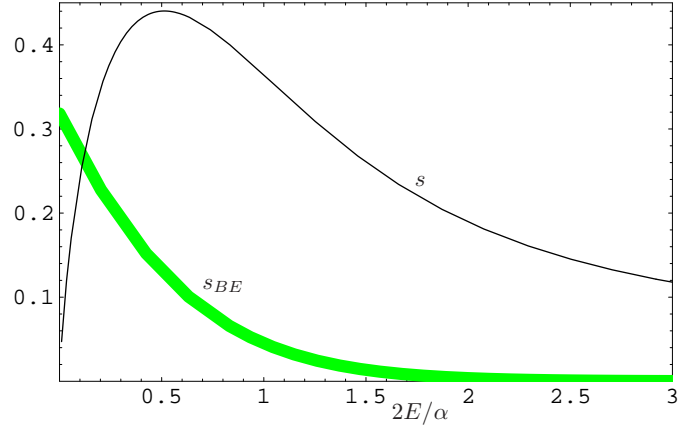}
\end{center}
\caption{The distribution functions $s_{BE}(2E/\alpha)$ and
 $s(2E/\alpha)$.
} \label{contour:fig1}
\end{figure}

We evaluate the response function using contour
integration. To do this we need to detour the contour so
that we exclude the singularity around $\tau=0$. Using the
contour integration result given in Appendix, we finally have
\begin{eqnarray}
S_+(\tau_0,E) &=&
\frac {\alpha} {2\pi}
s_{BE}(2E/\alpha)+O\left(\frac1{\kappa^2} \right)\,,
\label{s+} \\
\label{s-}
S_-(\tau_0,E)
&=& - \frac{\alpha}{2\pi}\, \frac{\alpha}{ 2\kappa}\left\{
\left[~\frac{\alpha}{2E}\left(9 \cosh (3\alpha
\tau_0)+2\cosh \alpha\tau_0\right)- \frac{2E}{\alpha}
\left(\cosh (3\alpha \tau_0 ) + 2\cosh (\alpha \tau_0
)\right)\right] s_{BE}(2E/\alpha)  \right.
\nonumber \\
&&\qquad\left.+\frac{1}{\pi} \left[3\sinh (3\alpha \tau_0)
+ 2\sinh(\alpha \tau_0)\right] s(2E/\alpha)
 \right\} +O\left(\frac1{\kappa^2} \right)
\,,
\end{eqnarray}
where
\begin{eqnarray} \label{sE}
s_{BE}(\zeta )&=&  \frac \zeta
{e^{\pi \zeta }-1}\,, \qquad\quad
s(\zeta) =
\int_{0}^{\infty} \frac{ dk}{k} \Big(\frac{\zeta |k-1|}{e^{\pi\zeta
|k-1|}-1} -\frac{\zeta |k+1|}{e^{\pi\zeta |k+1|}-1}
\Big) \,.
\end{eqnarray}

The distribution function
$s_{BE}(2E/\alpha)$
is the Bose-Einstein one,
finite at $E=0$ and
decays exponentially for large $E$.
On the other hand,
$s(2E/\alpha)$
shows no definite statistics but fuzziness,
vanishes at $E=0$, and decreases slowly as
$s(2E/\alpha)\sim \alpha/(4\pi^2E)$ for large $E$.
The behaviors of $s_{BE}$ and $s$ are plotted in
Fig.~\ref{contour:fig1}.

It is to be noted that
$S_+$ only reproduces the commutative result
and presents no $O(1/\kappa)$ correction.
Lorentz symmetry breaking appears in $S_-$ at $O(1/\kappa)$.
Here the correction term shows the preferred-frame effect,
the dependency of the detector time $\tau_0$.

Since the transition probability in Eq.~(\ref{exrate}) also
depends on how the detector couples to the complex scalar
field, one may tune the detector without modifying the
field theory of the massless complex scalar field. Thus if
one tune the detector so that $M$ is hermitian ($M_-=0$),
then the detector does not see the Lorentz violation at
$O(1/\kappa)$. Since $M_+(E,E_0)\geq |M_-(E,E_0)|$ and
$M_-$ is sensitive to the violation of the Lorentz
symmetry, one may consider the detector with
$M_+(E,E_0)=M_-(E,E_0)$, maximally sensitive to the
violation of Lorentz symmetry.

\begin{figure}[htbp]
\begin{center}
\includegraphics[width=.6\linewidth,origin=tl]{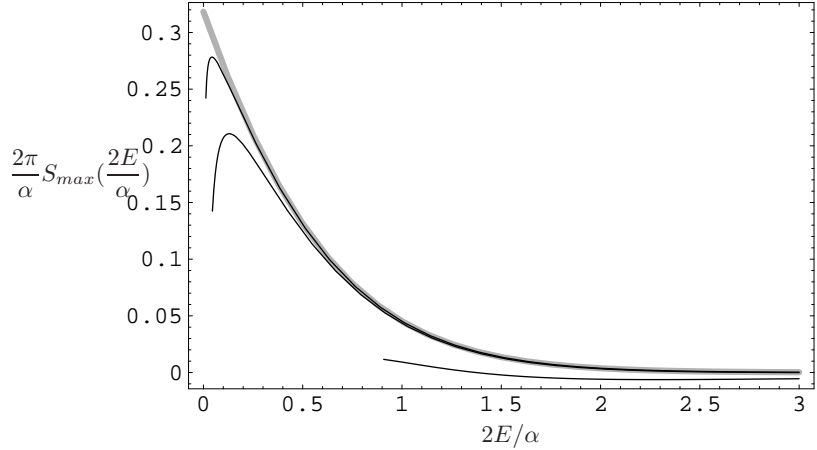}
\end{center}
\caption{Plot of the response function $S_{max}(\tau_0, E)$.
$\alpha$ is set as $5\cdot 10^{-5}\kappa$.
The shaded curve represents the commutative case,
the Bose-Einstein result.
The solid curves denote the response function at time
$\alpha\tau_0=0$, $\alpha \tau_0=1$, $\alpha\tau_0=2$,
respectively, from top to bottom.
At each time scale, the curves are shown for the valid
energy range
of the relevant perturbation result.} \label{contour:fig2}
\end{figure}

In this maximal case, the response function is given by
\begin{eqnarray} \label{S:max}
S_{max}(E,\tau_0)=S_+(E,\tau_0)+S_-(E,\tau_0) .
\end{eqnarray}
An explicit example of the response function with the
restrictions given in Eqs.~(\ref{valid range}) and (\ref{rest:2})
below is plotted in Fig.~\ref{contour:fig2}.
As seen in the figure,
the response function deviates slightly
from the Bose-Einstein result.

The Lorentz symmetry breaking results in $\tau_0$ dependence
and the excitation rate depends on the proper-time.
At $\tau_0 \sim 0$, $S_-$ is given by $s_{BE}$ term only
since $\sinh(\alpha \tau_0)=0$.
Since the $O(1/\kappa)$ correction term must be smaller
than the $O(\kappa^0)$ term,
we have the range of validity of the
results~(\ref{s+}) and (\ref{s-}),
\begin{eqnarray}
\frac{\alpha^2}{\kappa}  \ll E \ll \kappa \,.
\end{eqnarray}
The maximal response function becomes
\begin{eqnarray} \label{t=0}
S_{max}(\tau_0=0,E)
 &\simeq&\frac{E}{\pi}
    \frac{1}{e^{2\pi
    E/\alpha}-1}\left[1-\frac{\alpha}{2\kappa}\left(
        \frac{11\alpha}{2E}-
        \frac{6E}{\alpha}\right)\right].
    \nonumber
\end{eqnarray}

At $\alpha \tau_0 \sim O(1)$, the infrared part of
$S_-$ is dominated by $s_{BE}$ term and the ultraviolet part is
dominated by $s$ term.
For $1 \ll \alpha \tau_0 < \alpha \tau_c
= (\ln(2\kappa/(3\alpha)))/3 $, the proper time
contribution becomes large and thus the range of validity
is limited as
\begin{eqnarray} \label{valid range}
\frac{9\alpha}{2\kappa} e^{3\alpha \tau_0} \ll
\frac{2E}{\alpha} \ll \frac{2\kappa}{\alpha} e^{-3\alpha
\tau_0} \,.
\end{eqnarray}
(At $\tau_0 = \tau_c $  the left and right-hand side become equal
and the range is empty).
In the UV part of the spectrum, another restriction appears by comparing
the behavior of $s(2E/\alpha)$ and $s_{BE}(2E/\alpha)$:
\begin{eqnarray} \label{rest:2}
\left(\frac{\alpha}{2E}\right)^{2}\,e^{2\pi E/\alpha} \ll
\frac{4\pi^3\kappa}{3\alpha}
 e^{-3\alpha\tau_0}.
\end{eqnarray}

When $ \tau_0 \gg \tau_c $,
we cannot use the result of $W_\pm$
in Eqs.~(\ref{diff:Gpm}) and (\ref{diff:Gmm}).
This is because after this time, if $\tau\sim \tau_0 \to \infty$,
the four vector  $X(\tau)-X(\tau_0)$
is not time-like but becomes almost light-like.
Therefore, the series expansion used in
Appendix is not valid anymore.
In the limit, $|X(\tau)-X(\tau_0)| \rightarrow 0$ with
$X(\tau_0)\rightarrow \infty$, the coefficient $(a-b)$ in
the integral Eq.~(\ref{int:A1}) in Appendix  vanishes.
Therefore, the integration becomes a
Gaussian-like integral proportional to $\int_0^\infty d\mu
e^{-\kappa X \mu^2/2}$. After explicit evaluation,
the Wightman function $W_+(X(\tau),X(\tau_0))$ can be
shown to be
\begin{eqnarray} \label{W:lightlike}
W_+(X(\tau),X(\tau_0)) \propto \sqrt{\kappa}
[X(\tau)-X(\tau_0)]^{-{3/2}}=\sqrt{\kappa}
\left[\frac{2}{\alpha}\sinh \frac{\alpha (\tau-\tau_0)}{2}
\sinh \frac{ \alpha(\tau+\tau_0)}{2}\right]^{-3/2}.
\end{eqnarray}
In addition, for time-like interval of
$|X(\tau)-X(\tau_0)|$ when  $|\tau_0-\tau| \rightarrow \infty$,
$W_+(X(\tau),X(\tau_0))$ is
roughly of the form, $\displaystyle
-\frac{\alpha^2}{8\pi^2}\sinh^{-2}(\alpha
(\tau-\tau_0)/2)$.
Using these result, one may write
the long-time behavior of the response function  as
\begin{eqnarray} \label{S:lightlike}
S(\tau_0, E)\propto e^{-3\alpha \tau_0/2}\quad ~~~~\mbox{ when }
\tau_0 \gg \tau_c\, .
\end{eqnarray}

This implies that the uniformly accelerating
detector is not excited by the interaction of the
scalar field in $\kappa-$Minkowski vacuum
when $\tau_0 \gg \tau_c$.
This result can be understood as follows: The Wightman
function measures the correlation of the field between the
two points. If the detector is uniformly accelerating, and
if two points are connected by almost light-like path, the
energy difference between two points become very large even
for the case of small separation (or momentum difference),
because of the dispersion relation~(\ref{Casimir}) between
the momentum and energy. The
difference of the oscillation frequency between the two
points may be very large even if the momentum difference of
the two is small.
Therefore, the correlation between
the two points decays very fast, due to the fast
oscillation (difference) of the energy eigenfunction.

This indicates that the $\kappa-$deformation completely
alters the vacuum structure of uniformly accelerating
observers. In the simplest model in $\kappa-$ Minkowski
spacetime, the (fuzzy) thermal spectrum of the Minkowski vacuum
is seen to a uniformly accelerating observer
for a long but finite duration of proper time,
$\tau_c \sim (\ln (\kappa/\alpha))\alpha$.
After this time, the response function
exponentially decays to zero.

\section{summary}

We have studied a uniformly accelerating particle detector, interacting with massless complex scalar field in $\kappa-$Minkowski spacetime.
Introducing the simplest interaction of the detector
with the scalar field we calculated the response function
using the first order perturbation theory.
The correction term depends on the proper time of the detector
as a result of the Lorentz symmetry breaking effect.

One may devise a maximally sensitive detector whose
response function gradually deviates from the Bose-Einstein
result as the proper time increases. This deviation of the
response function is much enhanced when the proper time
approaches to a critical time $\tau_c \sim
(\log\kappa/\alpha)/\alpha $. This critical time is of the
order of logarithmic value of Plank mass after which the
temperature of the Minkowski vacuum state seen by the
observer decreases to zero.

It may have some implication to blackhole radiation problem.
However small the spacetime noncommutativity is,
it can nullify the thermal spectrum of the blackhole eventually.
This implies that we need to examine
the blackhole radiation problem carefully if there is a
spacetime non-commutativity with $\kappa-$deformation.
Note, however, that the present result is not a conclusive
one since we have used the first order perturbation theory
in obtaining the response function. To have better
understanding on this feature one must incorporate higher
order perturbations or try non-perturbative calculations.
In addition, one may has to incorporate the coordinates of the
accelerating observer having the symmetry consistent
with the spacetime non-commutativity,
$[\hat t,\hat x^i]=i\hat x^i/\kappa$.
This will be considered in future publications.

\begin{acknowledgments}
This work was supported by the Korea Research Foundation
Grant funded by Korea Government(MOEHRD, Basic Research
Promotion Fund)" (KRF-2005-075-C00009;H.-C.K.) and in part
by Korea Science and Engineering Foundation Grant No.
R01-2004-000-10526-0
and through the the Center for Quantum Spacetime(CQUeST) of Sogang
University with grant number R11-2005-021.
\end{acknowledgments}

\begin{appendix}
\section{integration}

The integrations relevant for the calculation of the
response function are
\begin{eqnarray} \label{int:A1}
I(a,b) =  \int_0^1 dz~ e^{i (\log(1-z) \, a  + \, z
b) }
\\
J(a,b) =  \int_0^1 dz~ e^{i (\log(1+z) \, a + \, z b)}.
\end{eqnarray}
Putting $\log(1-z)=-t$, we can write $I(a,b)$ as
\begin{equation}
I(a,b) = \int_0^\infty dt ~e^{-t - i (t a - b ( 1-
e^{-t}))}\,.
\end{equation}

The integration has the main contribution around $t\sim 0$.
Thus we may expand $( 1- e^{-t})$ around $t=0$, if $a-b \ne
0$:
\begin{eqnarray}
I(a,b) &=& \int_0^\infty dt ~e^{-t - i t( a - b) }
\times e^{ ib( -\frac{t^2}2 \cdots )}\\
&=& \frac1 {1 + i (a-b)}
 - \frac{i b} {(1 +i(a-b))^3 } +  \cdots .
 \nonumber
\end{eqnarray}
If $a$ and $b$ are order of $\kappa$, we may rescale $a$
and $b$ to  $\kappa a$ and $\kappa b$, and expand the
result in $1/\kappa$ as,
\begin{eqnarray}
I(\kappa a,\kappa b)
 &=& -\frac{i}{\kappa(a-b)}+\frac{a}{\kappa^2(a-b)^3}+\cdots .
 \nonumber
\end{eqnarray}
Likewise for $J(a,b)$ we have
\begin{eqnarray}
J(a,b) &=&\int_0^1 dz~
 e^{i ((z- \frac{z^2}2 +\cdots) \, a + \, z b)}
\\ \nn
&=& \frac{e^{i(a+b)}-1}{i (a+b)}
- \frac{a\,e^{i(a+b)}}{2(a+b)}
-i\frac{a\, e^{i(a+b)}}{(a+b)^2}
+\frac{a\,(e^{i(a+b)}-1)}{(a+b)^3}+\cdots \,.
\end{eqnarray}
Rescaling $a$ and $b$ to  $\kappa a$ and $\kappa b$, we
have
\begin{eqnarray}
J(\kappa a,\kappa b)
&=& \frac{i}{ \kappa (a+b)}
- \frac{a}{\kappa^2(a+b)^3}+\cdots \,.
\end{eqnarray}

Finally, we list some integral formulas needed in Sec. III:
\begin{eqnarray}
\int_{-\infty}^{\infty} d\tau~ \frac{ e^{-i \zeta \tau} }
{\sinh^2(\tau -i\epsilon)} &=& -2\pi \zeta
\sum_{n=1}^\infty e^{-i\zeta \tau_n} =-\frac{2\pi
\zeta}{e^{\pi \zeta  }-1} \, ,
\\
\int_{-\infty}^{\infty}   d\tau~
\frac{ e^{-i\zeta\tau} \cosh (k \tau)}
{\sinh^3 (\tau -i\epsilon) }
&=& -\pi i  ( (k^2-1) -\zeta^2)
 \sum_{n=1}^\infty e^{-i\zeta\tau_n}
=-\frac{\pi i  ( (k^2-1) -\zeta^2) } {e^{\pi \zeta }-1} \,
 ,\\
\int_{-\infty}^{\infty} d\tau~ \frac{ e^{-i\zeta \tau}\,
\varepsilon(\tau) } {\sinh^2(\tau)} &=& \frac1{2\pi
i}\int_{-\infty}^{\infty} dk \int_{-\infty}^{\infty} d\tau~
\frac{ e^{-i\zeta \tau} } {\sinh^2(\tau )} \Big(
\frac{e^{ik\tau}-e^{-ik\tau}}{k-i\epsilon} \Big)
\\ \nn &=&
i \int_{-\infty}^{\infty}
\frac{ dk}{k}
\Big(\frac{|k-\zeta|}{e^{\pi|k-\zeta|}-1}
-\frac{|k+\zeta|}{e^{\pi|k+\zeta|}-1} \Big)
\\ \nn &=&
2i \int_{0}^{\infty} \frac{ dk}{k}
\Big(\frac{|k-\zeta|}{e^{\pi|k-\zeta|}-1}
-\frac{|k+\zeta|}{e^{\pi|k+\zeta|}-1} \Big) .
\end{eqnarray}

\end{appendix}

\vfill\eject


\end{document}